\documentstyle[12pt]{article}
\newcommand{\beq}{\begin{equation}}
\newcommand{\eeq}{\end{equation}}
\newcommand{\beqa}{\begin{eqnarray}}
\newcommand{\eeqa}{\end{eqnarray}}
\newcommand{\ba}{\begin{array}}
\newcommand{\ea}{\end{array}}
\newcommand{\CR}{\nonumber \\}
\newcommand{\pa}{\partial}
\newcommand{\A}{\alpha}
\newcommand{\B}{\beta}

\newcommand{\G}{\gamma}

\newcommand{\tP}{\widetilde{\Pi}}

\newcommand{\oO}{{\cal O}}

\newcommand{\am}{a_{\{ m\}}}

\newcommand{\bpm}{b'_{\{ m\}}}
\renewcommand{\thefootnote}{\fnsymbol{footnote}}

\begin{document}
\begin{titlepage}
\begin{flushright}
hep-th/9807183 \\
YITP-98-44 \\
July, 1998
\end{flushright}
\vspace{0.5cm}
\begin{center}
{\Large \bf 
Seiberg-Witten Theory as $d<1$ Topological Strings}
\lineskip .75em
\vskip2.5cm
{\large Katsushi Ito}, {\large Chuan-Sheng Xiong}
\vskip 1.5em
{\large\it Yukawa Institute
for Theoretical Physics \\  Kyoto University, Kyoto 606-8502, Japan}  
\vskip 3.5em
{\large Sung-Kil Yang}
\vskip 1.5em
{\large\it Institute of Physics, University of Tsukuba \\
Ibaraki 305-8571, Japan}
\end{center}
\vskip3cm
\begin{abstract}
In view of two-dimensional topological gravity coupled to matter, 
we study the Seiberg-Witten theory for the low-energy behavior of $N=2$
supersymmetric Yang-Mills theory with $ADE$ gauge groups. We construct a new
solution of the Picard-Fuchs equations obeyed by the Seiberg-Witten
periods. Our solution is expressed as the linear sum over the infinite
set of one-point functions of gravitational descendants in $d<1$ topological
strings. It turns out that our solution provides the power series expansion
around the origin of the quantum moduli space of the Coulomb branch. 
For $SU(N)$ gauge group we show how the Seiberg-Witten periods are 
reconstructed from the present solution.
\end{abstract}
\end{titlepage}
\baselineskip=0.7cm

\newpage
\renewcommand{\thefootnote}{\arabic{footnote}}

Recently it has become clear that the Seiberg-Witten (SW) solution \cite{SeWi}
of $N=2$ supersymmetric Yang-Mills theory shares several common properties
with two-dimensional topological field theory. For $ADE$ gauge groups
this relation is recognized if one considers the SW solution in view of
the $ADE$ singularity theory. The relevance of the $ADE$ singularity theory
to the SW solution was speculated when the original $SU(2)$ SW solution
was generalized to the case of other gauge groups \cite{Le}. 

To further develop the idea it was crucial that the $ADE$ singularity,
which is usually expressed in terms of three complex variables, admits the
description just by using a single variable \cite{LW,EY}. As a result
the SW curve is regarded as the fibration over ${\bf CP}^1$ whose fiber
is the single-variable version of the superpotential for the $ADE$
topological Landau-Ginzburg (LG) models in two dimensions.

An important consequence is then that the Picard-Fuchs equations obeyed
by the SW period integrals are shown to be equivalent to the Gauss-Manin
system for the $ADE$ topological LG models plus in addition the scaling
equation \cite{ItYa1}.
Note that the Gauss-Manin system, on the other hand, is known to yield
the topological recursion relation \cite{WDW} when topological LG 
models are coupled to two-dimensional topological gravity \cite{EYY}.
Another interesting topological field theoretic aspect of the SW theory is
that the holomorphic prepotential of $N=2$ Yang-Mills theory satisfies the
WDVV equations \cite{MaMiMo}. 
A simple proof based on topological LG models is given
in \cite{ItYa3}.

Thus we have seen that fundamental properties of two-dimensional topological
field theory such as LG superpotentials, the Gauss-Manin system and
the WDVV equations are all exhibited by the SW solution of four-dimensional
$N=2$ Yang-Mills theory. One may then ask if there is any room in the SW theory
where two-dimensional topological gravity plays a role. Our purpose in this
article is to present explicitly the affirmative answer to this question.
As will be shown, the effect of two-dimensional topological gravity can be
captured when we study in detail the behavior of the SW period integrals
near the origin of the quantum moduli space of the Coulomb branch.

For $ADE$ type gauge group $G$ with rank $r$, the underlying Riemann 
surface \cite{MaWa} to
describe the low energy behavior of the Coulomb branch of $N=2$
Yang-Mills theory is given by 
\begin{equation}
W_{G}(x; t^{0}, \cdots, t^{r-1})-z-{\mu^{2}\over z}=0,
\label{eq:spe}
\end{equation}
where $W_{G}(x; t^{0}, \ldots, t^{r-1})$ is identified as the superpotential 
for the LG models of type $G$ with flat coordinates
$t^{\A}$ $(\A=0,\cdots, r-1)$.\footnote
{Note that the flat coordinates are labeled
differently from \cite{ItYa1}. Here we follow the convention customary in
two-dimensional topological gravity.}
The overall degree of $W_G$ is equal to $h$, the Coxeter number of $G$, and
$\mu^2=\Lambda^{2h}/4$ with $\Lambda$ being the dynamical scale. $t^{\A}$ 
has the degree $r_{\A}=h-e_{\A+1}+1$ where $e_{i}$ is the $i$-th exponent 
of $G$ $(e_1=1,\, e_r=h-1)$.
The SW differential 
\begin{equation}
\lambda_{SW}={x\over 2\pi i }{dz \over z}
\label{SWdiff}
\end{equation}
is used to define the period integrals
\begin{equation}
a^{I}=\oint_{A_{I}}\lambda_{SW}, \hskip10mm 
a_{D}^{I}=\oint_{B_{I}}\lambda_{SW}, \hskip10mm  I=1, \cdots , r
\label{periods}
\end{equation}
where $A_{I}$ and $B_{I}$ are canonical one-cycles on the curve.

Employing the technique of topological LG models, one can 
obtain the Picard-Fuchs equations for 
the period integrals $\Pi=(a^{I}, a_{D}^{I})$ \cite{ItYa1}. 
For example, from the operator product expansions for the primary fields 
$\oO_{\A}=\partial_{t^{\A}}W(x)$ with
$\partial_{t^{\A}}={\partial\over \partial t^{\A}}$:
\begin{equation}
\oO_{\A}(x)\oO_{\B}(x)=\sum_{\G=0}^{r-1}{C_{\A\B}}^{\G}(t)\oO_{\G}(x)
+Q_{\A\B}(x)
\partial_{x}W(x),
\end{equation} 
where $\partial_{x}Q_{\A\B}=\partial_{t^{\A}}\partial_{t^{\B}}W$,
one gets the Gauss-Manin system
\begin{equation}
\partial_{t^{\A}}\partial_{t^{\B}}\Pi
-\sum_{\G=0}^{r-1} {C_{\A\B}}^{\G}(t)
\partial_{t^{\G}}\partial_{t^{0}}\Pi=0.
\label{eq:gauss}
\end{equation}
Similarly, from the scaling relation for the LG superpotential:
\beq
x\partial_x W+\sum_{\A=0}^{r-1}
 r_{\A} t^\A{\partial W \over \partial t^\A} =hW,
\eeq
one obtains 
\begin{equation}
\left(\sum_{\A=0}^{r-1} r_{\A} t^{\A}\partial_{t^{\A}}+h\mu\partial_{\mu}
-1\right)\Pi=0.
\label{eq:scale}
\end{equation}
The remaining equation is obtained by regarding the LHS of the 
spectral curve (\ref{eq:spe}) as the superpotential of 
topological ${\bf CP}^1$ model \cite{Du}.
Namely, from the superpotential 
\begin{equation}
W_{{\bf CP}^{1}}=z+{\mu^{2}\over z}-t^{0},
\end{equation}
one may obtain the differential equation
\begin{equation}
\left( (\mu\pa_{\mu})^2 -4 \mu^{2} \pa_{t^{0}}^{2}\right) \Pi=0.
\label{eq:cp1}
\end{equation}
Note that the $\log \mu^{2}$ and $t^{0}$ play a role of flat
coordinates
of the topological ${\bf CP}^1$ model.
The Picard-Fuchs  equations (\ref{eq:gauss}), (\ref{eq:scale}) and 
(\ref{eq:cp1}) completely characterize the structure of the periods.
The weak-coupling ($\mu\sim 0$)  analysis shows that the prepotential
agrees with the microscopic calculations up to one-instanton level
\cite{KLT,inst}.

We now discuss the topological gravity coupled to the $ADE$ 
topological minimal model at genus zero \cite{WDW,DVV}. 
The topological minimal model associated with the $ADE$
type Lie group $G$ with rank $r$  
is obtained by twisting the $N=2$ minimal models with the central
charge $c=3d$ with $d=(h-2)/h$. The primary field 
$\oO_{\A}$ $(\A=0,\cdots, r-1)$ is a  BRST invariant observable, which has 
the ghost number charge $q_{\A}=(e_{\A+1}-1)/h$.\footnote{Notice that
the ghost number charge is equal to the degree divided by $h$.}
In particular, $\oO_{0}$ is the identity operator.

Coupling to topological gravity, we obtain gravitational descendants
$\sigma_{n}(\oO_{\A})$ ($n=0,1,2,\cdots$) of $\oO_{\A}$. Here 
we define $\sigma_{0}(\oO_{\A})=\oO_{\A}$.
Note that the identity operator $\oO_{0}$ and its first descendant 
$\sigma_{1}(\oO_{0})$ are identified with the puncture operator $P$
and the dilaton operator, respectively.
To each $\sigma_{n}(\oO_{\A})$ we may associate a coupling constant
$t_{n}^{\A}$.
We refer to the space spanned by  $t_{n}^{\A}$ as the large phase space.
The small phase space is defined by putting
$t_{n}^{\A}=0$ $(n\geq 1)$ except $t^{0}_{1}=-1$ and $t_{0}^{\A}\neq 0$.
$t_{0}^{\A}$is actually  the flat coordinate $t^{\A}$.

Then the correlation functions are given by the derivatives of the
free energy $F_{0}[t]$:
\begin{equation}
\langle \prod_{i}\sigma_{n_{i}}(\oO_{\A_{i}})\rangle
=\left(\prod_{i}\partial_{t_{n_{i}}^{\A_{i}}}\right)F_{0}[t].
\end{equation}
In particular, in the small phase space, we have 
\begin{equation}
\langle P \oO_{\A} \oO_{\B} \rangle=\eta_{\A\B},
\end{equation}
where  $\eta_{\A\B}=\delta_{e_{\A+1}+e_{\B+1},h}$ is the flat metric.
Structure constants ${C_{\A\B}}^{\G}(t)$ are given by the three
point functions:
\begin{equation}
\langle\oO_{\A}\oO_{\B}\oO^{\G}\rangle={C_{\A\B}}^{\G}(t),
\end{equation}
where $\oO^{\G}=\eta^{\G\B}\oO_{\B}$ and 
$\eta^{\A\B}$ 
is the inverse matrix of $\eta_{\A\B}$.
In the large phase space, one can show that
$F_{0}[t]$ 
obeys the following equations\cite{WDW}:
\begin{itemize}
\item the dilaton equation
\begin{equation}
2 F_{0}[t]=\sum_{n,\A} t_{n}^{\A}\pa_{t_{n}^{\A}}F_{0}[t].
\label{eq:dil}
\end{equation}
\item
the ghost-number conservation equation (or $L_{0}$-constraint)
\begin{equation}
\sum_{n,\A}(n+b_{\A})  t_{n}^{\A}\pa_{t_{n}^{\A}}F_{0}[t]=0,
\label{eq:gh}
\end{equation}
where $b_{\A}=q_{\A}-(d-1)/2=e_{\A+1}/h$.
\item 
the topological recursion relation
\begin{equation}
\langle \sigma_{n}(\oO_{\A}) X Y \rangle = \sum_{\beta}
\langle \sigma_{n-1}(\oO_{\A})\oO_{\B}\rangle \langle \oO^{\B} X Y \rangle.
\label{eq:rec}
\end{equation}
\end{itemize}
In the small phase space, we have the puncture equations 
\begin{equation}
\langle P \prod_{i=1}^{s} \sigma_{n_{i}}(\oO_{\A_{i}})\rangle
=\sum_{i=1}^{s} \langle \prod_{j=1}^{s}\sigma_{n_{j}-\delta_{ji}}
 (\oO_{\A_{j}})\rangle.
\label{eq:punct}
\end{equation}

{}From the topological recursion relation (\ref{eq:rec}) and the puncture 
equation (\ref{eq:punct}), it is easy to show 
\begin{equation}
\langle\oO_{\A}\oO_{\B}
\sigma_{n}(\oO_{\G})\rangle
=\sum_{\B'}\langle \oO_{\A}\oO_{\B}\oO^{\B'}\rangle 
\langle P \oO_{\B'} \sigma_{n}(\oO_{\G})\rangle.
\end{equation}
This is equivalent to the differential equation
\begin{equation}
\left(\partial_{t^{\A}}\partial_{t^{\B}}
-\sum_{\B'=0}^{r-1} {C_{\A\B}}^{\B'}(t)
\partial_{t^{\B'}}\partial_{t^{0}}\right)\langle\sigma_{n}(\oO_{\G})\rangle=0,
\end{equation}
which is 
nothing but the Gauss-Manin system (\ref{eq:gauss})\cite{EYY}.
Then one can introduce the power series 
\begin{equation}
\tP_{\A}=\sum_{n\geq 0} c_{n,\A} \langle \sigma_{n}(\oO_{\A})\rangle
\mu^{\rho (n+\theta_{\A})},
\end{equation}
which satisfies the Gauss-Manin system.
Here  $c_{n,\A}$, $\rho$ and $\theta_{\A}$ are constants which are
determined by requiring that $\tP_{\A}$ obeys the scaling relation
(\ref{eq:scale}) and the ${\bf CP}^{1}$ relation (\ref{eq:cp1}). 
{}From the dilaton equation (\ref{eq:dil}) and the ghost number
conservation (\ref{eq:gh}) in the small phase space, 
we obtain
\begin{eqnarray}
&& \langle \sigma_{m}(\oO_{\B})\rangle
+\langle \sigma_{1}(P) \sigma_{m}(\oO_{\B})\rangle
-\sum_{\A} t^{\A}\langle \oO_{\A} \sigma_{m}(\oO_{\B})\rangle=0, \\
&& (m+b_{\B}) \langle \sigma_{m}(\oO_{\B})\rangle
-(b_{0}+1) \langle \sigma_{1}(P) \sigma_{m}(\oO_{\B})\rangle
+\sum_{\A} b_{\A} t^{\A} \langle \oO_{\A} \sigma_{m}(\oO_{\B})\rangle=0,
\end{eqnarray}
respectively. 
Eliminating the term $\langle \sigma_{1}(P) \sigma_{m}(\oO_{\B})\rangle$
in the above two equations, we get\footnote{Since $1-q_\alpha =r_\alpha/h$,
eq.(\ref{degsigma}) indicates that the degree of 
$\langle \sigma_{n}(\oO_{\B})\rangle$ equals $h(n+1+b_\beta)+1$. This indeed
agrees with the degree we can directly read off from the integral
representation (\ref{eq:onep}).}
\begin{equation}
\sum_{\A=0}^{r-1}(q_{\A}-1)t^{\A}\pa_{t^{\A}}
 \langle \sigma_{n}(\oO_{\B})\rangle
+(n+b_{\B}+b_{0}+1)  \langle \sigma_{n}(\oO_{\B})\rangle=0.
\label{degsigma}
\end{equation}
{}From this relation and (\ref{eq:scale})
we find $\rho=-1$ and $\theta_{\A}=b_{\A}+1$. We next check the 
${\bf CP}^{1}$ relation. From the puncture equation (\ref{eq:punct}), we have
\begin{equation}
 \langle P^{2} \sigma_{n}(\oO_{\A})\rangle
 = \langle \sigma_{n-2}(\oO_{\A})\rangle.
\end{equation}
Thus the ${\bf CP}^{1}$ relation yields the recursion relation for
$c_{n,\A}$:
\begin{equation}
c_{n+2,\A}={1\over 4} (n+b_{\A}+1)^2 c_{n,\A}.
\end{equation}
We need two extra terms 
$4 c_{-2,\A}\eta_{0,\A}\mu^{1-b_{\A}}/(1-b_{\A})^2+4 c_{-1,\A}
\eta_{\A\B}t^{\B}\mu^{-b_{\A}}$ 
in addition to $\tP_{\A}$ in order to satisfy the ${\bf CP}^{1}$ relation.
To summarize the solution to the Picard-Fuchs equations is given by
\begin{equation}
\Pi_{\A}=A_{\A} 
\sum_{n\geq 0} {\Gamma(n+{b_{\A}\over2})^2 \over
\Gamma(1+{b_{\A}\over 2})^2} \langle \sigma_{2n-1}(\oO_{\A})\rangle
\mu^{-(2n+b_{\A})}
+B_{\A} 
\sum_{n\geq -1} {\Gamma(n+{b_{\A}+1\over2})^2 \over
\Gamma({b_{\A}+1\over 2})^2} \langle \sigma_{2n}(\oO_{\A})\rangle
\mu^{-(2n+b_{\A}+1)},
\label{eq:sol}
\end{equation}
where
$\langle \sigma_{-1}(\oO_{\A})\rangle\equiv \langle P \oO_{\A}\rangle
=\eta_{\A\B}t^{\B}$, 
$\langle \sigma_{-2}(\oO_{\A})\rangle\equiv \langle P^{2}
\oO_{\A}\rangle=\eta_{0\A}$, $\Gamma (x)$ is the Gamma function and
$A_{\A}$, $B_{\B}$ are certain constants. 

We next investigate the physical meaning of the solutions (\ref{eq:sol}) 
of the Picard-Fuchs equations.
The SW period integrals $\Pi=(a^{I}, a_{D}^{I})$ should be expressed in
terms of $\Pi_{\A}$. Obviously, the weak coupling solutions do not 
correspond to the present ones  because 
they are the expansion around $\mu=0$ and exhibit the logarithmic behavior.
The solutions (\ref{eq:sol}) are, on the other hand,
expansions around $\mu=\infty$. 
Hence we expect that these correspond to the period integrals expanded around
the origin of the moduli space $t^{\A}=0$. 
At this point, one cannot expect any massless soliton, and hence there appear
no logarithmic solutions. 

In the following we check this claim in the case of $A_{N-1}$ type
gauge groups explicitly.
Firstly we examine the simplest case, i.e. $A_{1}$ case, in which the
superpotential is given by $W_{A_{1}}=x^{2}-u$. 
The periods $(a, a_{D})$ are evaluated by  the elliptic
integral.
The solution at the weak coupling region is given in terms of the
hypergeometric function \cite{KLT}
\begin{eqnarray}
a(u)&=&{1\over \sqrt{2}} \sqrt{u}
F({1\over4},{1\over4};1;{\Lambda^{4}\over u^{2}}), \CR
a_{D}(u)&=& -{i\sqrt{u} \over \sqrt{2}\pi }
\left\{ 
F({1\over4},{1\over4};1;{\Lambda^{4}\over u^{2}})
 \log \left({\Lambda^{4}\over u^{2}}\right)\right. \CR
&& \left. +\sum_{n=1}^{\infty} 
{({1\over4})_{n} (-{1\over4})_{n}\over (1)_{n}^2}
\left({\Lambda^{4}\over u^{2}}\right)^{2}
\mbox{[}
-2\psi(n+1)+\psi(n+{1\over4})+\psi(n-{1\over4})\mbox{]}
\right\},
\label{eq:su2sol}
\end{eqnarray}
where $\Lambda^{4}=\mu^{2}/4$, $(x)_{n}\equiv \Gamma(x+n)/\Gamma(x)$ and 
$\psi(x)={d\log\Gamma(x)\over dx}$.
The hypergeometric function $F(a,b;c;z)$ is defined by 
\begin{equation}
F(a,b;c;z)=\sum_{n=0}^{\infty} {(a)_{n} (b)_{n}\over (c)_{n}}
{z^{n}\over n!}.
\end{equation}
In order to make analytic continuation to the region around $u=0$, we
use the formulas for the hypergeometric functions \cite{Erd};
\begin{eqnarray}
F(a,b;c;z) =& & {\Gamma(b-a)\Gamma(c)\over \Gamma(c-a)\Gamma(b)}
(e^{\pi i}z)^{-a} F(a,a-c+1;a-b+1;z^{-1}) \CR+
&& {\Gamma(a-b)\Gamma(c)\over \Gamma(c-b)\Gamma(a)}
(e^{\pi i}z)^{-b} F(b,b-c+1;b-a+1;z^{-1}), 
\label{eq:analy1}
\end{eqnarray}
\begin{eqnarray}
{\Gamma(a)\over\Gamma(c)} F(a,a;c;z)=&&
{(e^{-\pi i}z)^{-a}\over \Gamma(c-a)}
\sum_{n=0}^{\infty}
{(a)_{n} (1-c+a)_{n}\over(1)_{n}^{2}} z^{-n} 
[ \log(e^{-\pi i}z)+h_{n}], \label{eq:analy2} \\
h_{n}=&&2\psi(1+n)-\psi(a+n)-\psi(1-c+a+n)-\pi\cot\pi(c-a), \nonumber
\end{eqnarray}
where in (\ref{eq:analy2}) we take the branch $|\arg(e^{-\pi i}z)|<\pi$.
Applying (\ref{eq:analy1}) and (\ref{eq:analy2}) to the solutions
(\ref{eq:su2sol}), we obtain
\begin{eqnarray}
a(u) &&\!\!\!\!\!\!\!\!\!
={\sqrt{u}\Gamma({1\over2})\over 4\sqrt{2}}
\left(e^{\pi i}{\Lambda^{4}\over u^{2}}\right)^{{1\over4}}
\left\{
\Gamma(-{1\over 4})^{2}
F(-{1\over4},-{1\over4};{1\over2}; {u^{2}\over \Lambda^{4}})
+2\left(e^{\pi i}{\Lambda^{4}\over u^{2}}\right)^{-{1\over2}}
\Gamma({1\over 4})^{2} 
F({1\over4},{1\over4};{3\over2}; {u^{2}\over \Lambda^{4}})
\right\}, \CR
a_{D}(u)&&\!\!\!\!\!\!\!\!
=-{i\Lambda\over 8\pi \Gamma({1\over2})}
\left\{
\Gamma(-{1\over 4})^{2}
F(-{1\over4},-{1\over4};{1\over2}; {u^{2}\over \Lambda^{4}})
-2\left({u\over \Lambda^{2}}\right)
\Gamma({1\over 4})^{2} 
F({1\over4},{1\over4};{3\over2}; {u^{2}\over \Lambda^{4}})
\right\}.
\label{eq:pera1}
\end{eqnarray}

Let us consider the topological gravity. In this case, the observables are 
the gravitational descendants $\sigma_{n}(P)$ of the puncture operator $P$.
The one-point function $\langle \sigma_{n}(P)\rangle$
is given by
\begin{equation}
\langle \sigma_{n}(P)\rangle={(-u)^{n+2}\over (n+2)!}. 
\end{equation}
Thus the formula $\Pi_{\A}$ for $\A=0$ in (\ref{eq:sol}) may be
evaluated explicitly. We get 
\begin{eqnarray}
\sum_{n\geq 0} {\Gamma(k+{b_{0}\over2})^2 \over
\Gamma(1+{b_{0}\over 2})^2} \langle \sigma_{2n-1}(P)\rangle
\mu^{-(2n+b_{0})}
&=& 16 {-u\over \mu^{1\over2}}
F({1\over4},{1\over4};{3\over2}; {u^{2}\over 4\mu^{2}}), \CR
\sum_{n\geq -1} {\Gamma(k+{b_{0}+1\over2})^2 \over
\Gamma({b_{0}+1\over 2})^2} \langle \sigma_{2n}(P)\rangle
\mu^{-(2n+b_{0}+1)}&=& 16 \mu^{1\over2} 
F(-{1\over4},-{1\over4};{1\over2}; {u^{2}\over 4\mu^{2}}),
\end{eqnarray}
where $b_{0}={1\over2}$. 
Let us define $I_{0}(u,\mu)$ by choosing 
$A_{0}=\Gamma(1+b_{0}/2)^2/\Gamma(b_{0})$
and $B_{0}=\Gamma((1+b_{0})/2)^2/\Gamma(b_{0})$ for $\Pi_{0}$ 
in (\ref{eq:sol});
\begin{equation}
I_{0}(u,\mu)={1\over \Gamma({1\over2})}
\mu^{{1\over2}}
\left\{
\Gamma(-{1\over 4})^{2}
F(-{1\over4},-{1\over4};{1\over2}; {u^{2}\over 4\mu^{2}})
-{u\over \mu}
\Gamma({1\over 4})^{2} 
F({1\over4},{1\over4};{3\over2}; {u^{2}\over 4\mu^{2}})
\right\}.
\label{eq:sola1}
\end{equation}
$I_{0}(u,\mu)$ and $I_{0}(u,-\mu)$ are two independent solutions to the
Picard-Fuchs equation for $SU(2)$ gauge group.
Comparing (\ref{eq:pera1}) with (\ref{eq:sola1}), we obtain
\begin{eqnarray}
a &=& {1\over 8\sqrt{2}\pi} (I_{0}(u,\mu)-I_{0}(u,-\mu)), \CR
a_{D} &=& -{i\sqrt{2}\over8\pi} I_{0}(u,\mu).
\end{eqnarray}

We now generalize the $A_{1}$ result to the $A_{N-1}$ case.
We consider the topological gravity coupled with $A_{N-1}$-type
topological minimal models \cite{DVV}.
Let $W(x)=x^{N}-\sum_{i=2}^{N} s_{i} x^{N-i}$ be the LG
superpotential of $A_{N-1}$ type.
The one-point function of the $n$-th gravitational descendant 
$\sigma_{n}(\oO_{\A})$ of a primary field $\oO_{\A}$ is given by \cite{EYY}
\begin{equation}
\langle \sigma_{n}(\oO_{\A}) \rangle
={\Gamma(b_{\A})\over \Gamma(b_{\A}+n+2)}
\oint_{x=\infty} {d x\over 2\pi i}W(x)^{n+1+b_{\A}}, \hskip10mm   n\geq -2,
\label{eq:onep}
\end{equation}
where $b_{\A}={\A+1\over N}$.
Expanding $W(x)^{n+1+b_{\A}}$ around $x=\infty$, we have
\begin{equation}
W(x)^{n+1+b_{\A}}
=x^{N(n+1+b_{\A})}\sum_{m_2\geq 0, \cdots m_{N}\geq 0}
{(-1)^{\sum_{i=2}^{N} m_{i}}
\Gamma(n+2+b_{\A}) \over \Gamma (n+2+b_{\A}-\sum_{i=2}^{N} m_{i})}
 \prod_{i=2}^{N} {s_{i}^{m_{i}}\over m_{i}!} x^{-\sum_{i=2}^{N} i m_{i}}.
\end{equation}
Taking the residue at $x=\infty$, the non-zero contribution in the sum
(\ref{eq:onep}) occurs for the case $\sum_{i=2}^{N} i m_{i}=N(n+1)+\A+2$.
{}Since $m_{N}=n+1+{\A+2-\sum_{i=2}^{N-1}i m_{i}\over N}$, the
one-point 
function becomes 
\begin{equation}
\langle \sigma_{n}(\oO_{\A}) \rangle
= \sum_{\{ m_{i}\} \atop m_{N}=n+1+b_{\A}-\bpm: 
\mbox{\scriptsize integer}}
{\Gamma(b_{\A}) \over \Gamma(1-\am+\bpm)}(-1)^{\am+m_{N}}
\left(\prod_{i=2}^{N-1} {s_{i}^{m_{i}}\over m_{i}!}\right) 
{s_{N}^{m_{N}}\over m_{N}!},
\label{eq:onep1}
\end{equation}
where $\am=\sum_{i=2}^{N-2} m_{i}$, $\bpm=(\sum_{i=2}^{N-2} i
m_{i}-1)/N$ and $\{ m_{i} \}$ denotes $m_2\geq 0, \cdots, m_{N-1}\geq 0$.
As in the case of $A_{1}$, we define  $I_{\A}(s_{i}, \mu)$ by choosing
$A_{\A}=\Gamma(1+b_{\A}/2)^2/\Gamma(b_{\A})$ and
$B_{\A}=\Gamma((1+b_{\A})/2)^2/\Gamma(b_{\A})$ in (\ref{eq:sol});
\begin{equation}
I_{\A}(s_{i}, \mu)={1\over \Gamma(b_{\A})}
\sum_{n\geq 0} \Gamma({n+b_{\A}-1\over 2})^{2}
\langle \sigma_{n-2} (\oO_{\A})\rangle \mu^{-(n+b_{\A}-1)}.
\label{eq:i1}
\end{equation}
One may regard $I_{\A}(s_{i},\mu)$ and  $I_{\A}(s_{i},-\mu)$ as
independent solutions of the Picard-Fuchs equations.
Combining (\ref{eq:onep1}) and (\ref{eq:i1}), 
we obtain
\begin{eqnarray}
I_{\A}(s_{i}, \mu) &=& \sum_{\{ m_{i}\} \atop
N \bpm \equiv \A+1 (\mbox{\scriptsize mod} N)}
{(-1)^{\am}\over \Gamma(1-\am+\bpm)} 
\left(\prod_{i=2}^{N-1} {s_{i}^{m_{i}}\over m_{i}!}\right) \mu^{-\bpm} 
\CR
&& \!\!\!\!\!\!\!\!\!\!\!\!\!\!\!\!\!\!\!\!\!\!\!\!\!\!\!\!\!\!\!\!
 \times \left\{ \Gamma({\bpm\over2})^{2} 
F({\bpm\over2},{\bpm\over2};{1\over2};{s_{N}^2\over 4\mu^{2}})
-\Gamma({\bpm+1\over2})^{2} {s_{N}\over \mu}
F({\bpm+1\over2},{\bpm+1\over2};{3\over2};{s_{N}^2\over 4\mu^{2}})
\right\}. \CR
\end{eqnarray}

We next calculate the period integral of the SW differential 
around the origin  $s_{i}=0$ in the quantum moduli space. 
We make analytic continuation from the weak coupling region 
$s_{2}=\cdots=s_{N-1}=0$ and $s_{N}=\infty$
to the region around $s_{2}=\cdots=s_{N}=0$.
To evaluate the period integral in the weak coupling region, we will
use the formulas obtained by Masuda-Suzuki \cite{MaSu}.
{}From the Barnes type integral formulas, the period integrals 
$(a^{k},a^{k}_{D})$ $(k=1,\cdots, N$),
\footnote{For the $A_{1}$ case ($N=2$), we have $a^{1}=-\sqrt{2}a(u)$ and
$a_{D}^{1}=-\sqrt{{\pi\over2}} a_{D}(u)$.}
which are subject to the constraint
$\sum_{k=1}^{N}a^{k}=\sum_{k=1}^{N}a_{D}^{k}=0$, are given by 
\begin{eqnarray}
a^{k} &=& {s_{N}^{1/N}\over N}
\sum_{\{ m_{i}\}} e^{-2\pi i k \bpm} 
{\Gamma(\am-\bpm)\over \Gamma(1-\bpm)} \left(\prod_{i=2}^{N-1}
{\A_{i}^{m_{i}}\over m_{i}!}\right)
F({\bpm\over2},{\bpm+1\over2};1;{\Lambda^{2N}\over s_{N}^{2}}),  \CR
a_{D}^{k} &=& {s_{N}^{1/N}\over \pi i N}
\sum_{\{ m_{i}\}} e^{-2\pi i k \bpm} 
\Gamma(\am-\bpm)
\left({-2^{\bpm} \sin\pi \bpm \over 4 \pi^{3/2}}\right) \left(\prod_{i=2}^{N-1}
{\A_{i}^{m_{i}}\over m_{i}!}\right) \CR
&& \times \sum_{n=1}^{\infty} 
{\Gamma(n+{\bpm\over2})\Gamma(n+{\bpm+1\over2})\over \Gamma(n+1)^{2}}
\left({\Lambda^{2N}\over s_{N}^{2}}\right)^{n} \CR
&& \!\!\!\!\!\!\!\!\!\!\!\!\!\!\!\!
\times \left\{
-2\psi(n+1)+\psi(n+{\bpm\over2})+\psi(n+{\bpm+1\over2})
+\log \left({\Lambda^{2N}\over s_{N}^{2}}\right)+2\pi\cot \pi\bpm
\right\}, \CR
\end{eqnarray}
where $\A_{i}=s_{i}/s_{N}^{i/N}$ and $4\mu^{2}=\Lambda^{2N}$.
By using the formula (\ref{eq:analy1}), $a^{k}$ becomes
\begin{eqnarray}
a^{k} &=& {1\over N}
\sum_{\{ m_{i}\}}
e^{-2\pi i k \bpm} \Gamma(\am-\bpm)
\left( \sum_{i=2}^{N-1}
{s_{i}^{m_{i}}\over m_{i}!} \right) \mu^{-\bpm} e^{-{\pi i\bpm\over2}}
{\sin \pi\bpm \over \pi} \CR
& & \times {1\over \pi}
\left\{
\sin{\pi\bpm\over 2} \Gamma({\bpm\over 2})^2
F({\bpm\over2},{\bpm\over2};{1\over2};{s_{N}^2\over 4\mu^{2}})
\right. \CR
&& \left. +i \cos {\pi\bpm\over 2} \Gamma({\bpm+1\over2})^{2} {s_{N}\over \mu}
F({\bpm+1\over2},{\bpm+1\over2};{3\over2};{s_{N}^2\over 4\mu^{2}})
\right\}.
\end{eqnarray}
This turns out to be 
\begin{equation}
a^{k}=-{1\over 4\pi i N}\sum_{\A} e^{-2\pi i k b_{\A}}
\left(
I_{\A}(s_{i}, \mu)-I_{\A}(s_{i}, -\mu)
\right) .
\end{equation}
For $a_{D}^{k}$, applying the formula (\ref{eq:analy1}) and (\ref{eq:analy2}),
we get 
\begin{eqnarray}
a_{D}^{k} &=& {1\over \pi i N}
\sum_{\{ m_{i}\} }
e^{-2\pi i k \bpm} \Gamma(\am-\bpm) 
\left( \sum_{i=2}^{N-1}
{s_{i}^{m_{i}}\over m_{i}!} \right) \pi (-1)^{{1\over2}-\bpm}\mu^{-\bpm}\CR
&& 
\times \left\{
\sin^{2}{\pi\bpm\over 2} \Gamma({\bpm\over 2})^2
F({\bpm\over2},{\bpm\over2};{1\over2};{s_{N}^2\over 4\mu^{2}})
\right. \CR
&& \left. - \cos^{2} {\pi\bpm\over 2} \Gamma({\bpm+1\over2})^{2}
 {s_{N}\over \mu}
F({\bpm+1\over2},{\bpm+1\over2};{3\over2};{s_{N}^2\over 4\mu^{2}})
\right\}.
\end{eqnarray}
This is shown to be
\begin{equation}
a_{D}^{k}={-1\over 4 N\sqrt{\pi}}
\sum_{\A}e^{-2\pi i k b_{\A}}
\left({e^{-\pi i b_{\A}}\over \sin\pi  b_{\A}}I_{\A}(s_{i}, \mu)
-\cot \pi b_{\A} I_{\A}(s_{i}, -\mu)\right) .
\end{equation}
So far we have calculated the period integrals for $A_{N-1}$ case. 
It would be a straightforward task to generalize the present results
to $D$-type gauge groups.

In the present work, we have seen that the period integrals of the 
Seiberg-Witten differential at the origin of the quantum moduli space
are described by two-dimensional topological gravity coupled to
topological $ADE$ minimal models. How do we then interpret the dependence on
$\mu$? It is tempting to regard (\ref{eq:spe}) as the superpotential for the
$ADE$ minimal models combined with the topological ${\bf CP}^1$ model.
In view of the ${\bf CP}^1$ model the power of $\mu^2$ counts the
degree of holomorphic maps from the Riemann surface to ${\bf CP}^1$,
and hence has to be non-negative integers. It is thus curious to have
negative fractional powers of $\mu^2$ for $n \geq 1$ in (\ref{eq:sol}).
On the other hand, there is another interpretation of (\ref{eq:spe}) in
terms of topological sigma models. According to \cite{OV}, the LG models
with superpotentials
\beq
W(x,y)=W_G(x;\, 0)-\rho y^{-h}
\label{deform}
\eeq
correspond to sigma models with the ALE target space of $ADE$-type
singularities;\\
$W_G(x;\, 0)=\rho$. Therefore, setting $z=y^h$, we may think of
the superpotential (\ref{eq:spe}) as deformations of (\ref{deform}).
The LG formulation of the sigma models on the ALE space describes the
Type IIB string on $K3 \times {\bf R}^6$ when the $K3$ surface is near the
$ADE$ singularities. Note here that Type IIB on ALE is equivalent to
Type IIA on the NS fivebrane \cite{OV}. Hence, identifying the Riemann
surface (\ref{eq:spe}) with the superpotential in the LG description of
deformed ALE sigma models seems consistent with the M-theory fivebrane
interpretation of the SW solution \cite{fivebrane}. Connections
between $ADE$ $d<1$ topological strings and the ALE sigma models have been
discussed on the basis of superconformal theories \cite{OV}. Clarifying the
relation between \cite{OV} and our present observation of SW theory
as $d<1$ topological strings will be important for deeper understanding.

There are several issues to be considered further in this direction. 
We have investigated the two-dimensional theory at genus zero and 
in the  small phase space.
Hence it is interesting to generalize the present approach to the case of
arbitrary genus and general background. 
After appropriate change of normalization of $t_{n}^{\A}$,
the power series solutions $I_{\A}(s_{i},\pm \mu)$ correspond to
loop operators in two-dimensional theory and are 
realized by $Z_{N}$ twisted bosons. 
The Virasoro or $W$-constraints in two-dimensional theory impose 
the non-trivial conditions on the prepotential in four dimensions.
In the four-dimensional sense, these generalizations might correspond to
including the higher derivative terms or the coupling to four-dimensional
gravity. It is also interesting to investigate the relation between the present
formalism and the approach based on the Whitham hierarchy \cite{MaWa, Wh}.


\vskip3mm\noindent
The work of S.K.Y. was supported in part by Grant-in-Aid for Scientific
Research from the Ministry of Education, Science and Culture
(No. 09640335). S.K.Y. would like to thank A. Tsuchiya for stimulating
discussions.

\newpage

\end{document}